# Sub-exponential complexity of regular linear *CNF* formulas

Bernd R. Schuh

Dr. Bernd Schuh, D-50968 Köln, Germany; bernd.schuh@netcologne.de

**Abstract.**

The study of regular linear conjunctive normal form (*LCNF*) formulas is of interest because exact satisfiability (XSAT) is known to be NP-complete for this class of formulas. In a recent paper it was shown that the subclass of regular <u>exact</u> *LCNF* formulas (*XLCNF*) is of sub-exponential complexity, i.e. XSAT can be determined in sub-exponential time, namely $O(n^{\sqrt{n}})$. Here I show that this class is just a subset of a larger class of *LCNF* formulas which display this very kind of complexity. To this end I introduce the property of *disjointedness* of *LCNF* formulas, measured, for a single clause $C \in F$, by the number of clauses which have no variable in common with *C*, $d_C$. If for a given *LCNF* formula *F* all clauses have the same disjointedness *d* we call *F d-disjointed* and denote the class of such formulas by *dLCNF*. *XLCNF* formulas correspond to the special case $d = 0$. One main result of the paper is that the class $(\leq d)LCNF_+^l$ of all monotone *l*-regular *LCNF* formulas which are *d'-disjointed,* with $d' \leq d$, is of sub-exponential complexity. This result can be generalized to show that all monotone, *l*-regular *LCNF* formulas *F* which have a bounded <u>mean disjointedness</u> $\overline{d}_F = \frac{1}{|F|}\sum_{C \in F} d_C$, are of sub-exponential XSAT-complexity, as well.

**Notations and definitions.**

Throughout this communication I will adopt the notation used in [1,2,3] which is shortly repeated here. A Boolean formula in conjunctive normal form (*CNF*) by definition is a conjunction of clauses, where each clause is a disjunction of literals. A literal is an occurrence of a Boolean variable or its negation. In a linear *CNF* formula any two clauses have at most one variable in common. The class of such formulas is denoted by *LCNF*. In an exact linear formula any two clauses have <u>exactly</u> one variable in common. The class is denoted by *XLCNF*. A monotone formula contains positive literals only. Monotony is denoted by a subscript +, e.g. *XLCNF$_+$*. *l*-regularity is the property that each variable occurs *l* times. It is denoted by a superscript *l*, e.g. *CNF$^l$*. Finally, uniformity means that every clause contains the same number of literals. If this number is *k*, one writes e.g. *k-CNF*.

XSAT is the problem of deciding whether for a given *CNF* formula *F* there is a truth assignment (also called model) that satisfies exactly one literal in each clause of *F*. If there is at least one such assignment *F* is said to be x-satisfiable, otherwise x-unsatisfiable. If not stated otherwise the following notation is used for *CNF* formulas: $m=|F|$ for the number of clauses of *F*, $n=|V(F)|$ for the number of variables of *F*, $l(x)=|\{C \in F : x \in V(C)\}|$ for the occurrence of variable x, $k_C=|V(C)|$ for the number of variables in clause $C$. The representation of a formula by its i*ncidence matrix* $f_{Cx}:=1$ iff $x \in V(C)$ and $f_{Cx}:=0$ otherwise, can be useful in proofs.

The following definitions are crucial for the main results.

*Definition 1*: For a clause *C* of a linear *CNF* formula *F* its *disjointedness* $d_C$ is defined as the number of clauses which have no variable in common with clause *C*, i.e. $d_C = |\{C' \in F : V(C) \cap V(C') = \emptyset\}|$. Clause pair members C, C' with $V(C) \cap V(C') \neq \emptyset$ will be called "connected".

*Definition 2*: A *LCNF* formula *F* is called *d-disjointed* iff all clauses of *F* have the same disjointedness $d \in N'$.

Note that exact *LCNF* formulas can now be addressed as zero-disjointed *LCNF*.
Furthermore the following abbreviations are useful:

*Definition 3a:* Mean clause length of a *CNF* formula *F*: $\overline{k_F} := \frac{1}{m} \sum_{C \in F} |C|$

*Definition* 3b: Mean squared clause length: $\overline{k_F^2} = \frac{1}{m} \sum_{C \in F} |C|^2$

*Definition* 3c: Mean occurrence: $\overline{l_F} = \frac{1}{n} \sum_{x \in V(F)} l(x)$

Definition 3d: Mean squared occurrence: $\overline{l_F^2} = \frac{1}{n} \sum_{x \in V(F)} l(x)^2$

*Definition 4*: Mean disjointedness of a *LCNF* formula *F*: $\overline{d_F} := \frac{1}{m} \sum_{C \in F} d_C$ .

We also introduce the notion of i*ndependence* of two variables via

*Definition 5*: Let *F* be a linear CNF formula. Two variables $x, x' \in V(F)$ are called *independent* iff they have no clause in common, i.e. if $\{C \in F : x \in V(C) \wedge x' \in V(C)\} = \emptyset$ .

*Definition* 6: For a variable $x \in V(F)$ its i*ndependence* $v_x$ is defined as the number of variables independent of *x,* i.e. $v_x = |\{x' \in V(F)\} \setminus \{x\} \setminus \{x' \in V(F) : \exists C \in F : x \in V(C) \wedge x' \in V(C)\}|$ .

Finally we define the mean independence by

*Definition 7*: Mean independence of a formula *F*: $\bar{v} := \frac{1}{n} \sum_{x \in V(F)} v_x$ .

**Motivation**.

In a recent paper it was shown that monotone regular *XLCNF* formulas are of sub-exponential complexity, i.e. their decidability with respect to XSAT (exact satisfiability) is of order $O(n^{\sqrt{n}})$ , [3]. This result is of interest because XSAT was identified as NP-complete for monotone *LCNF* and this result was extended to *l*-regular *LCNF* and to X*LCNF* (without monotony) by Porschen et al., see e.g. [1,2]. Whether these results can be maintained for regular and even uniform instances as well was left as a conjecture by these authors.

Thus the question arises: are there larger subsets of regular *LCNF* , i.e. other than *XLCNF*, with sub-exponential complexity? In [5] this question has been studied for some simple subclasses of *k*-uniform *LCNF*. We now extend the study to proper *LCNF* formulas in general. Monotony and regularity are kept but the assumption of exactness is dropped. What difference does this make? In [3] it was shown that regularity forces uniformity in *XLCNF* formulas, and as a consequence the formula size depends only on the two parameters *k* (uniformity) and *l* (regularity):
$m = |F| = 1 + k(l-1)$ and $n = |V(F)| = km/l$ , to be precise. This is no longer the case if exactness is dropped. Then formula size depends on a third parameter which can be identified as the disjointedness of clauses. This additional degree of freedom in principal obstructs the method of proof used to show that complexity of XSAT is of order $O(n^{\sqrt{n}})$ in these formulas. The method uses the fact that the number of XSAT models, i. e. the number of satisfying assignments which evaluate exactly one literal per clause to "true", is bounded from above by the binomial coefficient $\binom{n}{m/l}$ , a fact that follows from straightforward considerations, [3]. (For a generalization of this formula to non-monotonous formulas see [4].) If formula size only depends on *k* and *l*, large formulas require large *k* (for fixed regularity *l*), and since $n = O(k^2)$ whereas $m/l = O(k) = O(\sqrt{n})$ , one gets the stated result. In the presence of non-zero disjointedness, however, one can construct arbitrarily large formulas even if all clause lengths are bounded from above by allowing for arbitrarily large disjointedness. We therefore introduce a new class of *LCNF*, characterized by disjointedness *d*, abbreviated $dLCNF$ and prove that the class of *l*-regular monotone $dLCNF_+^l$ is of complexity $O(n^{\sqrt{n}})$ with respect to XSAT (theorem 1). An enlargement of the subclass of $LCNF_+^l$ with this property is formulated in theorems 2 and 3.

**Structural properties of *LCNF* formulas**

As a first result we give a relation between formula size *m* and disjointedness $d_C$ for any clause *C* of the formula.

*Lemma* 1: If $F \in LCNF$, $C \in F$ is an arbitrary clause and $m = |F|$ the number of clauses, then

(1a) $\quad m = 1 + \sum_{x \in V(C)} l(x) - |C| + d_C$

Proof: For a linear formula the number of clauses connected to a given clause *C* is given by $\sum_{x \in V(C)} (l(x) - 1)$, because each variable $x \in V(C)$ of *C* occurs in *C* and in $l(x) - 1$ other clauses. These other clauses are pairwise different because of linearity. Thus the number of clauses <u>not</u> connected to *C*, i.e. $d_C$ by definition, is the total number of clauses *m* minus the clause itself, i.e. $m - 1$, minus $\sum_{x \in V(C)} (l(x) - 1)$. Solve for *m* to get the statement. ⊔

A different but instructive way to get the result is as follows. Choose a clause $C \in F$ and add all clauses C' with $V(C') \cap V(C) \neq 0$. Then one has a formula *F'* with $|F'| = 1 + \sum_{x \in V(C)} (l(x) - 1)$ many clauses. Any further clause $C'' \in F$ you can add to F' will have no variable in common with *C* because all occurrences of the variables $x \in V(C)$ are already present in *F'*. Thus

(1b) $\quad |F| - |F'| = d_C$

which is equivalent to (1a) and holds for each clause *C* of *F*.

A trivial consequence for <u>*l*-regular</u> *LCNF* is the

*Corrollary*: For $F \in LCNF^l$

(2) $\quad m = 1 + |C|(l-1) + d_C$

which follows immediately from (1a).

Since m-1 is constant for a given *F* one can deduce

*Lemma* 2: For $F \in LCNF^1$:

    *F* is *k*-uniform with some natural number *k* if and only if *F* is *d*-disjointed with some natural number *d* (including 0).

Proof: Follows from (2). If *F* is *k*-uniform, $|C|$ is constant. Then $d_C$ must be independent of *C*, and vice versa. □

Now disjointedness appears on the same footing as regularity and uniformity. In fact one can restate Lemma 2 as follows: <u>In regular *LCNF* uniformity forces disjointedness and vice versa.</u>

Before we proceed to the main theorem let us collect some properties of LCNF which in a certain sense are dual to the discussion above. Instead of focussing on the number of clauses we focus on the number of variables and write with the aid of definition 6:

*Lemma* 3: For $F \in LCNF$

$$n = 1 - l(x) + \sum_{C \in F} f_{Cx}|C| + v_x$$

Proof: This is the obvious analogue of lemma 1. For a given *x* collect all clauses *C* with $x \in V(C)$ and call the resulting formula $F_x$. Then $|V(F_x)| = \sum_{C \in F} f_{Cx}(|C|-1) + 1$, because: If $V(F_x)$ contained less elements then at least one *x'* other than *x* would be present in at least two clauses from $F_x$, in contradiction to linearity. Now $V(F) \setminus V(F_x)$ consists of variables which are independent of *x*. Therefore by definition 6 $v_x = |V(F) \setminus V(F_x)| = n - |V(F_x)| = n - 1 - \sum_{C \in F} f_{Cx}(|C|-1)$. Solve for n to get the stated result.

Since only the pure structure of the formula is concerned one can get the result by simply repeating the proof of lemma 1 for the "dual" problem obtained by interchanging clauses and variables, occurrence and clause length, and *n* and *m*. The only thing to note is that no two variables can occur in more than one clause. But that is a consequence of linearity.

Quite generally, without assuming regularity or uniformity, we can state the following symmetric relations for linear CNF formulas:

Lemma 4: For $F \in LCNF$ and with definitions 3-7 the following equations hold:

(3a) $\quad m = 1 + \dfrac{\bar{k}}{\bar{l}}(\overline{l^2} - \bar{l}) + \bar{d}$

(3b) $\quad n = 1 + \dfrac{\bar{l}}{\bar{k}}(\overline{k^2} - \bar{k}) + \bar{v}$

where all subscripts *F* have been omitted for clarity.

Proof: Sum (1a) over all $C \in F$ and devide by *m*. Likewise sum the equation in lemma 3 over all $x \in V(F)$ and divide by *n*. In both cases use $\bar{k}m = \sum_{C \in F}|C| = \sum_{x \in V(F)} l(x) = \bar{l}n$ and thus $m/n = \bar{l}/\bar{k}$.

Corollary: For $F \in LCNF^l$

(3c) $\quad m = 1 + \bar{k}(l-1) + \bar{d}$

(3d) $\quad n = 1 + l(\overline{k^2}/\bar{k} - 1) + \bar{v}$

**Complexity results.**

We denote the class of *d*-disjointed linear *CNF* formulas by *dLCNF* in the following, and the class of all *l*-regular *dLCNF* by $dLCNF^l$, except for *d*=0, where we will keep the usual notation *XLCNF*. All formulas in this class are uniform due to lemma 2, i.e. for every $F \in dLCNF^l$ there is some natural number *k* such that $|C| = k$ for all $C \in F$, and vice versa: for every $F \in k - LCNF^l$ (i.e. all *k*-uniform, *l*-regular *LCNF* formulas) there is some natural (or zero) *d* with $d_C = d$ for all $C \in F$. Note, however, that neither $k - LCNF^l \supseteq dLCNF^l$ nor vice versa holds.

According to (3c) the four quantities *m, k, l* and *d* within the class $dLCNF^l$ are related by

(4) $\quad m = 1 + k(l-1) + d$.

This relation holds for formulas $F \in k - LCNF^l$ as well, but then *l* and *k* are considered to be given, and various values *d* of disjointedness are allowed, whereas for $F \in dLCNF^l$ *l* and *d* are considered to be fixed, and *k* can vary throughout the class. We note in passing that allowed *k*- and/or *d*-values are restricted by the condition that *m* and *n* must be integers. Since we are concerned with XSAT satisfiability also $m/l = 0 \bmod(l)$ is a necessary condition, because *LCNF* formulas violating this condition are x-unsatisfiable. Therefore $k = (1+d) \bmod(l)$ is a necessary condition which also guarantees $m, n \in N$.

We now turn to the question of XSAT complexity of the new subclasses of regular *LCNF* and prove as a preliminary

*Theorem* 1: For $F \in dLCNF_+^l$ XSAT is decidable in time $O(n^{\sqrt{n}})$ up to polynomial corrections, where $n = |V(F)|$.

*Proof*: If $m \neq 0 \bmod(l)$ $F$ is x-unsatisfiable, see e.g. theorem 4 in [3]. This condition is easily checked. If $m = 0 \bmod(l)$ the rest of the proof follows the lines of proof of theorem 6 in [3]. Here is a short summary. All assignments which are XSAT solutions are among the $\binom{n}{m/l}$ assignments which make the total number of true literals equal to *m*. For *l* and *d* fixed *m* and *n* can only grow with *k*. One can solve (4) for *k*, insert the result in $n = km/l$ and solve for *m/l* to get

(*) $\quad m/l = \dfrac{1+d}{2l}\left(1 + \sqrt{1 + \dfrac{4nl(l-1)}{(1+d)^2}}\right)$ .

From (*) one can deduce

(**) $\quad \sqrt{n(l-1)/l} \leq m/l \leq (1+\varepsilon)\sqrt{n(l-1)/l}$

for arbitrarily small $\varepsilon$ and sufficiently large *n*. Thus $m/l = O(\sqrt{n})$ for large n and *d* and *l* fixed. As a consequence $\binom{n}{m/l}$ is of order $O(n^{\sqrt{n}})$ with polynomial corrections at most. A sub-exponential algorithm to decide XSAT then consists of determining all $\binom{n}{m/l}$ assignments and test each of them (in polynomial time) for x-satisfiability. □

Theorem 1 is a natural generalization of theorem 6 in [3], which states that monotone *l*-regular X*LCNF* are of the very same type of sub-exponential complexity. Since exact linearity is just a special case of disjointedness, namely *d*=0, theorem 6 in [3] is a special case of theorem 1.
One can generalize the result by allowing disjointedness to take all possible values up to a certain constant, i.e. consider instances from $(\leq d')LCNF^l := \{F \in LCNF^l : \exists d \leq d' : F \in dLCNF^l\}$ . Thus for $d \leq d' : LCNF^l \supseteq (\leq d')LCNF^l \supseteq dLCNF^l$ , and in particular

$\quad (\leq d')LCNF^l \supseteq XLCNF^l$

for any natural number *d'* including zero. One can now show sub-exponential XSAT-complexity for arbitrarily larger subclasses of $LCNF_+^l$.

*Theorem* 2: For $F \in (\leq d')LCNF_+^l$, $d' \in \{0,1,2,...\}$ XSAT is decidable in time $O(n^{\sqrt{n}})$ up to polynomial corrections.

*Proof*: For any $d \leq d'$ one infers from (*) $m/l \leq \frac{1+d'}{2l}\left(1 + \sqrt{1 + \frac{4nl(l-1)}{(1+d')^2}}\right)$. Again $m/l = O(\sqrt{n})$ up to polynomial corrections at most, such that the reasoning in the proof of theorem 1 can be adopted in theorem 2. □

Note that the uniformity parameter *k* is the quantity that drives the size of a formula for bounded values of *d*, just as in the *XLCNF* case. The proof in theorems 1 and 2 (and likewise in theorem 3 to follow) does not work for *k*-uniform regular *LCNF*. The reason is that *d* is of order *n* in this case which follows from (4) and $km = ln$ :

$$1 + d = nl/k - k(l-1)$$

such that also $m/l = O(n)$ as is apparent from (*).

A different generalization is possible by dropping the disjointedness condition, and with it uniformity. To this end we make use of definitions 3 and 4. If the reference is clear the subscript *F* will be dropped.

According to the corollary to Lemma 4 we have

(5)    $m = 1 + \bar{k}(l-1) + \bar{d}$

We can now generalize theorem 2 to the class of regular *LCNF* formulas with bounded mean disjointedness : $(\bar{d} \leq D)LCNF^l := \{F \in LCNF^l : \bar{d}_F \leq D\}$ .

*Theorem* 3: XSAT is decidable in time $O(n^{\sqrt{n}})$ up to polynomial corrections, within the class $(\bar{d} \leq D)LCNF_+^l$ for any fixed natural number *D*.

*Proof*: If there is no uniformity and no overall disjointedness one gets $\sum_{C \in F}|C| = nl = m\bar{k}$ instead of $km = nl$. Together with equation (5) one can infer (*) with *d* replaced by $\bar{d}$. Since $\bar{d} \leq D$ for any *F* from the class $(\bar{d} \leq D)LCNF_+^l$, one has

$$m/l \leq \frac{1+D}{2l}\left(1+\sqrt{1+\frac{4nl(l-1)}{(1+D)^2}}\right)$$

and again $\binom{n}{m/l}$ is $O(n^{\sqrt{n}})$ at most. The rest of the proof can be transferred from theorem 1. □

**Conclusion.**

In [3] the sub-exponential XSAT decidability of monotone exact linear *CNF* was proven. This finding was extended to special subsets of k-uniform $LCNF_+^l$ in [5]. These results could be generalized to more classes of *LCNF* formulas by theorems 1-3 in this paper. The crucial property is the notion of disjointedness which may be interpreted as a measure of how tight the fabric of clauses and variables in a linear *CNF* formula is knit. The less connected the clauses of a linear *CNF* instance are, i. e. the larger the overall disjointedness *d* is, the more likely it appears that the instance escapes the fate of sub-exponential complexity. At least it escapes the method of proof used here to demonstrate sub-exponential complexity. One cannot exclude that indeed the whole class of monotone regular *LCNF* might turn out to be of this diminished complexity, possibly provable with different methods. It seems more probable that in addition to the numerous regular *LCNF* instances which have been identified by our theorems as being of restricted complexity there is a host of instances which are not. The theorems also give a hint on where to find them: at the far end of increasing disjointedness. To state the results in a somewhat sloppy way one might say that the NP-hard *LCNF* instances (if $NP \neq P$) are to be found in the regime of "loosely connected" regular *LCNF* with unbounded values of disjointedness. It is possible, however, to construct "loosely connected" regular *LCNF* instances which display sub-exponential complexity still, (see [5] for details).

**References**.

[1] T. Schmidt, *Computational complexity of SAT, XSAT and NAE-SAT for linear and mixed Horn CNF formulas*, Ph.D. thesis, Institut für Informatik, Univ. Köln, Germany (2010).

[2] S. Porschen; T. Schmidt, E. Speckenmeyer, A. Wotzlaw, *XSAT and NAE-SAT of linear CNF classes*, Discrete Appl. Math. 167 (2014) 1-14.

[3] B.R. Schuh, *Exact satisfiability* of *linear CNF formulas*, Discrete Appl. Math. 251 (2018) 1-4.

[4] B.R. Schuh, *Sub-exponential Upper Bound for #XSAT of some CNF Classes*, https://arxiv.org/abs/1803.07376.

[5] B.R. Schuh, *XSAT of Linear CNF Formulas*, https://arxiv.org/abs/1711.07474